\documentclass[journal]{IEEEtran}
\usepackage{amsmath,amssymb,amsfonts}
\usepackage{array}
\usepackage[caption=false,font=normalsize,labelfont=sf,textfont=sf]{subfig}
\usepackage{textcomp}
\usepackage{stfloats}
\usepackage{url}
\usepackage{verbatim}
\usepackage{graphicx}
\usepackage{subcaption}
\usepackage{cite}
\usepackage{caption}
\usepackage{algorithm}
\usepackage{algpseudocode}
\usepackage{tikz}
\usepackage{xcolor}
\usepackage{footnote}
\usepackage{multirow}
\usepackage{tabulary}
\usepackage{lipsum}
\usepackage{booktabs}
\usepackage{soul}
\usepackage{tablefootnote}
\usepackage{xcolor}
\usepackage{colortbl}
\usepackage[dvipsnames]{xcolor}
\usepackage{float} 

\begin{document}
\title{Permutation-Equivariant Learning for Dynamic Security Assessment of Power System Frequency Response \\
\thanks{This work was supported in part by the National Science Foundation (NSF) under Grant No. 2044629, and in part by CURENT, which is an NSF Engineering Research Center funded by NSF and the Department of Energy under NSF Award EEC-1041877.}
\thanks{Zelaya-Arrazabal, Martinez-Lizana, and Pulgar-Painemal are with the 
Department of Electrical Engineering and Computer Science, University of Tennessee, 
Knoxville, TN 37996, USA (e-mail: fzelayaa@vols.utk.edu, smartinezlizana@ieee.org, 
hpulgar@utk.edu). Jin Zhao is with Trinity College Dublin, Dublin, Ireland 
(e-mail: zhaoj6@tcd.ie).}}
\author{%
Francisco Zelaya-Arrazabal,~\IEEEmembership{Member,~IEEE,} 
Sebastian Martinez-Lizana,~\IEEEmembership{Graduate Student Member,~IEEE,} 
Héctor Pulgar-Painemal,~\IEEEmembership{Senior Member,~IEEE}
Jin Zhao,~\IEEEmembership{Member,~IEEE}
\vspace{-3em}} 

\maketitle
\begin{abstract}
This paper presents a hybrid model–AI framework for real-time dynamic security assessment of frequency stability in power systems. The proposed method rapidly estimates key frequency parameters under a dynamic set of disturbances, which are continuously updated based on operating conditions and unit commitment. To achieve this, the framework builds on a modal-based formulation of the system frequency response (SFR), which leverages the system’s eigenstructure to predict key frequency stability metrics. A Deep Sets-inspired neural network is employed to estimate the complex modal coefficients required by the modal-based SFR approach, formulated as a permutation-equivariant learning problem. This enables fast and accurate prediction of the frequency nadir and its timing across different operating conditions and disturbances. The framework achieves scalability by reusing precomputed modal structures and updating only the disturbance-specific coefficients. It demonstrates strong generalization capabilities without requiring an extensive set of operating scenarios during training or the widespread deployment of phasor measurement units (PMUs). The method is validated on the IEEE 39-bus and 118-bus systems, showing superior accuracy, robustness, and computational efficiency compared to purely data-driven approaches.
\end{abstract}
\begin{IEEEkeywords}
Frequency nadir, system frequency response, frequency stability, dynamic security assessment, deep Sets, permutation-equivariant.
\vspace{-0.2in}
\end{IEEEkeywords}
\section{Introduction}
\IEEEPARstart{G}{rowing} system variability, driven by increasing renewable integration, has elevated the importance of Dynamic Security Assessment (DSA) in modern power grids. DSA evaluates a power system’s ability to withstand a defined set of contingencies and transition to an acceptable steady-state condition \cite{sauer2007dynamic}. Depending on its timing and application, DSA is typically classified into offline and online approaches.

Offline DSA commonly relies on time-domain simulations or analytical methods applied across many operating scenarios to determine secure operating limits used by operators in real-time decision-making \cite{yu2005direct}. In contrast, online DSA leverages artificial intelligence (AI) or statistical techniques to rapidly assess system security following a disturbance, providing binary stability classification, stability indices, or margins \cite{de2023review,sun2007online,xu2012reliable}. The vast majority of recent research has focused on online DSA, which is inherently reactive—meaning that a contingency must occur before any assessment can be performed. This is often justified as a means to enhance situational awareness \cite{zhang2019method,wang2023neural,xie2021transfer,lin2019post,zhang2019prediction,wang2019integrating,zhang2022review,hossain2021graph,wang2020frequency}. However, from an operator’s perspective, this is not entirely satisfactory, as reactive indices typically offer only a few seconds for control action, which limits their practical value in critical decision-making.

Traditionally, DSA has focused on rotor angle and voltage stability \cite{sauer2007dynamic}. However, the rise of inverter-based resources and declining system inertia have made frequency stability a growing concern. Recent studies have incorporated frequency stability into the DSA framework by estimating system frequency response (SFR) or key parameters like frequency nadir, nadir timing, steady-state frequency, and RoCoF \cite{zhang2019method,wang2023neural,xie2021transfer,lin2019post,zhang2019prediction,wang2019integrating}. A growing trend in these studies is the application of AI techniques to estimate these parameters and support real-time operational decisions \cite{zhang2022review}. However, this trend presents several challenges. Purely data-driven models often require large volumes of historical data and diverse operating conditions to generalize well across unseen scenarios \cite{zhang2022review}. Moreover, these models must be capable of retraining online to adapt to evolving system conditions, which is computationally expensive and often impractical. To manage variability, graph-based and structure-aware models have been proposed to better capture system dynamics \cite{xie2021transfer,lin2019post,hossain2021graph}. While conceptually appealing, such approaches typically rely on high-dimensional input features derived from widespread real-time monitoring. Although technically feasible, deploying phasor measurement units (PMUs) at all nodes to support these models requires substantial infrastructure investment, making it economically unrealistic for most systems \cite{song2025steady}. Furthermore, PMU-based methods face additional challenges, including data loss, noise, communication delays, and burdensome preprocessing \cite{liu2023practical}—issues that become more pronounced when models require a vast number of measurements.

Therefore, there is a clear need for models that generalize across changing grid conditions without relying on extensive PMU deployment. These models should strike a balance between offline and online DSA: offering real-time evaluations without requiring a purely reactive approach. Instead, a predictive framework that leverages steady-state operating conditions to anticipate system behavior can enhance situational awareness and provide operators with more actionable insights and lead time for control. Such models are not meant to replace existing online DSA tools but to complement them by enhancing awareness of dynamic security margins and enabling more proactive decision-making.

This paper presents a hybrid model–AI framework for real-time dynamic security assessment of frequency stability. Building on the modal-based SFR and nadir prediction method from \cite{zelaya2025modal}, it uses a Deep Sets-inspired neural network to estimate modal coefficients, enabling fast and accurate prediction of frequency nadir and timing across diverse conditions. By replacing repeated analytical computations with data-driven estimation, the approach accelerates assessment while preserving physical interpretability. It strikes a balance between offline and online DSA by providing predictive insights without requiring a disturbance-triggered response. Validation on the IEEE 39-bus and 118-bus systems confirms its scalability and effectiveness. 

The main contributions of this work are: a) the development of a real-time dynamic assessment framework to enhance situational awareness of frequency stability; b) the design of a Deep Sets-inspired neural model that captures the permutation-equivariant nature of grid states and modal information; and c) a lightweight, predictive solution that operates from steady-state data without the need for widespread deployment of wide-area measurement systems.

This paper is structured as follows: Section II presents background on modal-based SFR and nadir estimation. Section III reviews the modularity of the modal-based approach using frequency control modes and modal coefficients. Section IV introduces the proposed frequency dynamic security assessment framework. Section V describes the estimation of complex modal weights using a Deep Sets-inspired neural network. Section VI evaluates the proposed framework on two benchmark systems. Finally, the conclusions of this work are presented in Section VII.

\section{Background on Modal-Based SFR and Nadir Estimation}

The modal-based approach for estimating the SFR and nadir enables fast and accurate assessment by leveraging the system's eigenstructure and frequency control modes. This method provides an explicit closed-form expression for the frequency trajectory following a power imbalance~\cite{zelaya2025modal}. 

Consider a power system described by a set of differential-algebraic equations $\dot{x} = f(x, y)$, $0 = g(x, y)$,
where $x \in \mathbb{R}^n$ is the vector of dynamic state variables and $y \in \mathbb{R}^l$ is the vector of algebraic variables. Linearizing the system around an equilibrium point $x_e$, yields,
\begin{equation}
    \Delta\dot{x} = A_s \Delta x
\end{equation}
where $A_s = J_1 - J_2 J_4^{-1} J_3$ is the reduced Jacobian obtained through Schur complement, and $\Delta x(0) = \Delta x_0 = x_0 - x_e$ denotes the initial deviation from the pre-disturbance state. The similarity transformation of $A_s$ is defined as $V^{-1}A_sV = W^\top A_s V = \Lambda$, where $V = \{v_i\} \in \mathbb{C}^{n \times n}$ and $W = \{w_i\} \in \mathbb{C}^{n \times n}$ are the right and left eigenvector matrices of $A_s$, respectively. The matrix $\Lambda = \text{diag} \{\lambda_i\} \in \mathbb{C}^{n \times n}$ contains the eigenvalues of $A_s$, and the index $i \in \mathcal{I} = \{1, 2, \dots, n\}$. The eigenvectors are assumed to be orthonormal, i.e., $\langle w_i, v_k \rangle = w_i^\top v_k = \delta_{ik}$ for all $i,k \in \mathcal{I}$, implying $W^\top = V^{-1}$. Using modal decomposition, the explicit solution for each state variable is given by:
\begin{equation}
    \Delta x_k(t) = \sum_{i \in \mathcal{I}} e^{\lambda_i t} \langle w_i, \Delta x_0 \rangle v_{i,k}, \quad \forall~ t \geq 0,
\end{equation}
where $v_{i,k}$ is the $k^{\text{th}}$ component of the $i^{\text{th}}$ right eigenvector.

The frequency at the center of inertia (coi) can be approximated as:
\begin{equation}
\label{eq:w_coi_approx}
    \Delta \omega_{coi}(t) \approx
    \sum_{z \in \mathcal{Z}} \sum_{m \in \mathcal{M}} C_z e^{\lambda_m t} \langle w_m, \Delta x_0 \rangle v_{m,z},
\end{equation}
where \( C_z = H_z / H_t \) is a weighting factor based on the generator inertia \( H_z \), with \( H_t = \sum_{z \in \mathcal{Z}} H_z \) as the total system inertia. The set $\mathcal{Z} \subset \mathcal{I}$ refers to the indices associated with generator speed states, while $\mathcal{M} \subset \mathcal{I}$ corresponds to the indices of frequency control modes, i.e. those modes that significantly influence the SFR.

Eq.~\eqref{eq:w_coi_approx} provides an explicit expression for the SFR as a function of time $t$ and the initial deviation $\Delta x_0 = x_0 - x_e$, where $x_e$ is the post-disturbance equilibrium point under a power imbalance $\Delta P_L$. Since $\Delta\omega_{coi}$ is primarily influenced by rotor speed and governor dynamics, it is sufficient to estimate only the corresponding subset of $\Delta x_0$. The steady-state frequency deviation resulting from a disturbance $\Delta P$ is given by~\cite{kundur2007power}:
\begin{equation}
    \Delta \omega=-\Delta P/\sum_{z \in \mathcal{Z}} \tfrac{1}{R_{z}}
    \label{eq:Post_freq}
\end{equation}
where $1/R_{z}$ is the speed regulation of generator $z$, and $\omega_e = \omega_0 + \Delta \omega$. Let $x_{\mathcal{S}}$ represent the state variables related to rotor speed and governor dynamics. Their equilibrium values $x_{\mathcal{S},e}$ are found by solving $0 = f_{\mathcal{S}}(x_{\mathcal{S},e}, y_e)$. For example, in a system with purely IEESGO model, the relevant variables are,
\begin{align}
    y_{1,i} &= y_{3,i} = -\frac{\omega_e}{R_{D,i}}, \\
    T_{m,i} &= P_{C,i} - \frac{\omega_e}{R_{D,i}}, \quad \forall i \in \{1, \dots, n_g\},
\end{align}
where $n_g$ is the number of generators. The full equilibrium vector $x_e$ is completed by setting all other entries to zero. The frequency nadir is obtained by differentiating Eq.~\eqref{eq:w_coi_approx} and solving for the time when the derivative equals zero. Due to the presence of multiple transcendental terms, a direct analytical solution is not feasible. To simplify, each term—associated with a frequency control mode—is approximated using a second-order Taylor expansion. Let \( \mu \) denote the number of frequency control modes, resulting in the following sum of polynomial expressions:

\begin{equation}
\label{eq:wcoi_der_general}
\frac{d \Delta\omega_{\text{coi}}}{dt} = \sum_{j=1}^\mu \left( \sum_{k=0}^m a_{jk} t^k \right) = \sum_{k=0}^m b_k t^k
\end{equation}
where \( a_{jk} \) is the coefficient of the \( k^{\text{th}} \) power of time from the \( j^{\text{th}} \) frequency control mode, and \( b_k = \sum_{j=1}^\mu a_{jk} \) is the aggregated coefficient for \( t^k \). The time at which the nadir occurs, \( t_{\text{nadir}} \), is obtained by solving the following equation:

\begin{equation}
\label{eq:tnadir_general}
\sum_{k=0}^m b_k t^k = 0, \quad t_{\text{nadir}} > 0
\end{equation}
Once \( t_{\text{nadir}} \) is determined from Eq.~\eqref{eq:tnadir_general}, the per unit frequency nadir can be computed by evaluating the expression for \( \Delta\omega_{\text{coi}}(t) \) at this instant. Specifically,

\begin{equation}
\label{eq:wnadir_general}
f_{\text{nadir}} = f_e + \Delta\omega_{\text{coi}}(t_{\text{nadir}})
\end{equation}
where \( f_e \) is the post-disturbance steady-state frequency, obtained from Eq.~\eqref{eq:Post_freq}. More details on how to obtain the derivative and each $a_{jk}$ coefficient can be found in \cite{}.

\section{Modular Parameterization of the SFR Using Modal Coefficients}

\subsection{Key parameters}
The closed-form expression for the SFR in Eq.~\eqref{eq:w_coi_approx} can be reformulated as a sum of complex-weighted exponential terms,
\begin{equation}
\label{eq:1conjugated_1real}
\Delta \omega_{\text{coi}}(t) \approx \sum_{i=1}^{n} e^{\lambda_i t} \gamma_i
\end{equation}
where \( \Lambda_\mathcal{M} = \{\lambda_1, \dots, \lambda_n\} \) and \( \Gamma_\mathcal{M} = \{\gamma_1, \dots, \gamma_n\} \) are the ordered sets of frequency control modes and their respective complex modal coefficients, indexed by \( \mathcal{M} \subset \mathcal{I} \). Each \( \gamma_i \) is computed as
\begin{equation}
\label{eq:modal_coefficients}
   \gamma_i = \sum_{z \in \mathcal{Z}} C_z \langle w_i, \Delta x_0 \rangle v_{i,z},
\end{equation}
where the contribution of each generator is weighted by its normalized inertia \( C_z \). The modes in \( \Lambda_\mathcal{M} \) are selected using the participation factor-guided methodology described in~\cite{}, and are primarily determined by the system operating condition. In contrast, the coefficients \( \gamma_i \in \Gamma_\mathcal{M} \) depend not only on the system state but also on the disturbance characteristics and unit commitment.

\subsection{Modular parametrization}
As presented in~\cite{zelaya2025modal}, sensitivity analyses indicate that \( \Lambda_\mathcal{M} \) remains largely invariant under changes in topology, load levels, and power flow, making it reusable across scenarios with the same unit commitment. Conversely, \( \Gamma_\mathcal{M} \) must be recomputed for each disturbance to accurately reflect the system's dynamic response. Taking into account this, let the system consist of \( n_g \) generators, and consider \( k \) disturbances, each producing a unique power imbalance \( \Delta P_L^{(j)} \). The corresponding set of complex modal coefficients, for all $k$ disturbances, can be grouped into the following collection,
\begin{align}
\label{eq:gamma_set}
\Gamma = \Big\{ \Gamma_\mathcal{M}^{(j)} = \Gamma_\mathcal{M}(&\text{modal structure}, \Delta x_0^{(j)}(\Delta P_L^{(j)}),\\
& \texttt{dist}^{(j)}, \mathbf{u}_{\text{on}}) \nonumber \Big|\, j = 1, 2, \dots, k \Big\}
\end{align}
where \( \mathbf{u}_{\text{on}} \in \{0,1\}^{n_g} \) is the binary unit commitment vector indicating the status (on/off) of each generator. The term \( \Delta x_0^{(j)} \) represents the initial state deviation induced by disturbance \( j \), and \( \texttt{dist}^{(j)} \) contains information about the disturbance type (e.g., generator outage or load imbalance). The modal structure comprises the dominant eigenvalues and the associated left and right eigenvectors \( W_\mathcal{M} \) and \( V_\mathcal{M} \), derived from the linearized system.

Once the modal structure \( \Lambda_\mathcal{M} \) is fixed for a given operating condition, the disturbance-specific sets \( \Gamma_\mathcal{M}^{(j)} \) must be computed for each of the \( k \) disturbances using Eq.~\eqref{eq:modal_coefficients}. These coefficients are then used in Eq.~\eqref{eq:1conjugated_1real} to evaluate the corresponding SFR trajectories and to compute the frequency nadir using Eqs.~\eqref{eq:wcoi_der_general}–\eqref{eq:wnadir_general}. This enables efficient analytical computation of key frequency response metrics across a wide range of disturbances using precomputed eigenstructure and scenario-specific inputs.


\section{Frequency Dynamic Security Assessment}
Online dynamic assessment requires fast estimation of the frequency nadir and its timing across multiple contingencies. The modal-based framework is well-suited for this task, offering a structured and efficient formulation of the system response. This section introduces a dynamic assessment approach that leverages the modularity of the formulation to enable scalable, high-speed evaluation of frequency stability under varying conditions.

\begin{figure}[b]
\centering
\includegraphics[width=0.95\columnwidth]{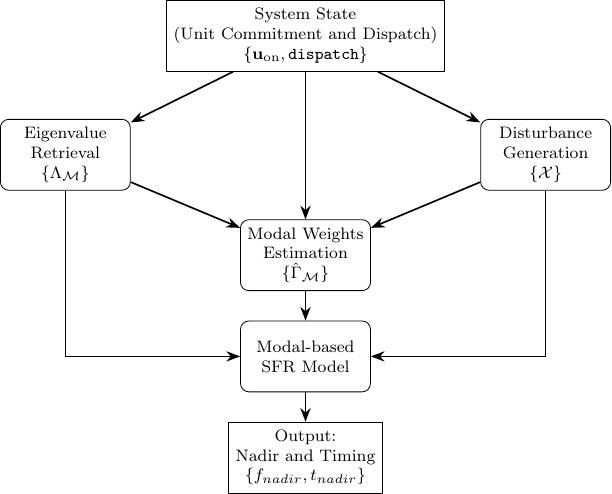}
\caption{SFR dynamic assessment framework}
\label{fig:sfr_framework}
\end{figure}
\subsection{General Framework}
The core idea behind SFR dynamic assessment is to take the current system conditions and evaluate a collection \( \Gamma \) corresponding to a list of potential disturbances. This list is adapted according to the system operating state and unit commitment. The final output of this framework is the frequency nadir and timing associated with each defined contingency. Figure~\ref{fig:sfr_framework} illustrates the overall architecture. This framework is composed of the following four subprocesses:

\subsection{Eigenvalue Retrieval}
Because dominant SFR eigenvalues are primarily influenced by the unit commitment, the \( \Lambda_\mathcal{M} \) set can be reused for all scenarios sharing the same commitment configuration. A bank of precomputed \( \Lambda_\mathcal{M} \) sets can be generated offline and stored. In real-time, the appropriate set is retrieved by matching the current unit commitment vector \( \mathbf{u}_{\text{on}} \in \{0,1\}^{n_g} \) with one in the database. This bank can be updated daily based on the unit commitment schedule determined by the day-ahead market, or alternatively, constructed from historical data that reflects typical operating conditions.

\subsection{Disturbance generation}
Based on the current system state, the framework automatically generates a set of possible disturbances that may lead to critical frequency deviations. Each disturbance is represented by a vector \( x_j = [\Delta P_L^{(j)}, \texttt{type}, \texttt{index}] \), where \( \Delta P_L^{(j)} \) denotes the magnitude of the power imbalance in megawatts, \texttt{type} is a binary indicator specifying the disturbance type (0 for generator trip, 1 for load disturbance), and \texttt{index} identifies the affected system element, such as a generator or load bus. For example, \( x_1 = [55, 0, 4] \) corresponds to a 55~MW trip at generator 4, while \( x_2 = [40, 1, 7] \) represents a 40~MW load disturbance at bus 7.

The complete set of candidate disturbances is defined as:
\begin{equation}    
\mathcal{X} = \left\{ x_j = \left[\Delta P_L^{(j)}, \texttt{type}, \texttt{index} \right] \,\middle|\, j = 1, 2, \dots, k \right\}
\end{equation}
The set \( \mathcal{X} \) is automatically generated within the framework based on real-time system conditions, including unit commitment, dispatch levels, and the identification of critical loads. Since the framework is intended to run periodically for situational awareness, \( \mathcal{X} \) is dynamically updated to reflect the current operating state.

This ensures adaptability: the list of possible disturbances reflects the current power injection from all generators when modeling generator trips, as well as critical load centers considered for load imbalance disturbances. 


\subsection{Complex Modal Weights}
This subprocess receives as input the unit commitment vector \( \mathbf{u}_{\text{on}} \in \{0,1\}^{n_g} \), the modal structure \( \Lambda_\mathcal{M} \), and the disturbance set \( \mathcal{X} \), and performs fast batch estimation of \( \widehat{\Gamma} = \left\{ \widehat{\Gamma}_\mathcal{M}^{(j)} \right\} \), where each set \( \widehat{\Gamma}_\mathcal{M}^{(j)} \) corresponds to a disturbance \( x_j \in \mathcal{X} \).

Estimation is carried out using a Deep Sets-inspired neural network—a deep learning model designed for permutation-equivariant inputs. In this model, the input—an unordered, variable-size set of information—is mapped directly to the estimated modal coefficient vectors \( \widehat{\Gamma}_\mathcal{M}^{(j)} \). This learning-based approach expedites dynamic analysis by bypassing the need to compute disturbance-specific initial conditions \( \Delta x_0 \) and each individual modal coefficient \( \gamma_i \) using Eq.~\eqref{eq:modal_coefficients} for every disturbance \( x_j \in \mathcal{X} \). The full model architecture and training methodology are described in the following section.

\subsection{Modal-based prediction}
Once the set of estimated modal coefficients \( \widehat{\Gamma} \) has been obtained for all disturbances in \( \mathcal{X} \), Eqs.~\eqref{eq:wcoi_der_general}–\eqref{eq:wnadir_general} can be applied to compute the frequency nadir and its timing. Alternatively, these values can be extracted directly from the reconstructed SFR response by evaluating Eq.~\eqref{eq:1conjugated_1real} over a defined time horizon. Once \( \Delta \omega_{\text{coi}}^{(j)}(t) \) is computed for each disturbance \( x_j \in \mathcal{X} \), the nadir magnitude and its corresponding time are defined as:
\begin{equation}
\Delta f_{\text{nadir}}^{(j)} = \min_{t \in [0, T]} \Delta \omega_{\text{coi}}^{(j)}(t), \quad
t_{\text{nadir}}^{(j)} = \arg\min_{t \in [0, T]} \Delta \omega_{\text{coi}}^{(j)}(t)
\end{equation}
where \( \Delta f_{\text{nadir}}^{(j)} \) denotes the frequency deviation from the post-disturbance steady-state condition \( x_e \). The actual nadir value is then computed as \( f_{\text{nadir}}^{(j)} = f_e + \Delta \omega_{\text{coi}}^{(j)}(t_{\text{nadir}}^{(j)}) \), where \( f_e \) is the post-disturbance steady-state frequency. The complete output of the dynamic assessment framework is therefore given by:

\begin{equation}   
\mathcal{Y} = \left\{ \left(f_{\text{nadir}}^{(j)}, t_{\text{nadir}}^{(j)} \right) \,\middle|\, j = 1, 2, \dots, k \right\}
\end{equation}

This modular structure enables near real-time predictions by retrieving frequency control modes and estimating their corresponding modal coefficients upfront. It supports rapid assessment of multiple contingencies without requiring full nonlinear simulations.

\section{Complex Modal Weights Estimation}
This section describes the estimation of the complex modal weights collection \( \widehat{\Gamma} = \left\{ \widehat{\Gamma}_\mathcal{M}^{(j)} \right\} \), which approximates the mapping defined in Eq.~\eqref{eq:gamma_set} using a deep learning model. The estimated mapping is given by:
\begin{equation}
\widehat{\Gamma}_\mathcal{M}^{(j)} = f\left( \Lambda_\mathcal{M}, \mathbf{u}_{\text{on}}, x_j \right)
\end{equation}

A key modeling challenge is that \( \Lambda_\mathcal{M} \) is a set of eigenvalues whose order may vary with the unit commitment. As a result, the model must operate on sets of eigenvalues rather than fixed-order feature vectors, and the output mapping must respect the input permutation. Formally, a function \( f \) is \textit{permutation invariant} if it satisfies \( f(X) = f(\pi(X)) \) for any permutation \( \pi \) of the input set \( X \), meaning the output remains unchanged regardless of input order. In contrast, a function is \textit{permutation equivariant} if \( f(\pi(X)) = \pi(f(X)) \), implying that the output changes in a manner consistent with the permutation of the input. This property is essential in our case, where, as shown in Eq.~\eqref{eq:1conjugated_1real}, the relationship between the \( i \)-th eigenvalue \( \lambda_i \) and its corresponding modal coefficient \( \gamma_i \) must be preserved, regardless of ordering.

Therefore, the estimation the set \( \widehat{\Gamma}_\mathcal{M}^{(j)} \) requires an architecture that respects permutation equivariance over the set \( \Lambda_\mathcal{M} \), while simultaneously be conditioned by \( \mathbf{u}_{\text{on}} \) and \( x_j \). 
\subsection{Deep Sets}
The Deep Sets framework provides a principled approach for learning functions over sets. As shown in~\cite{zaheer2017deep}, any permutation-invariant function \( \mathcal{F} \) defined over a set \( \mathcal{S} = \{ \mathbf{s}_1, \mathbf{s}_2, \dots, \mathbf{s}_n \} \) can be expressed as:
\begin{equation}
\label{eq:DeepSetModel}
\mathcal{F}(\mathcal{S}) = \Psi\left( \sum_{i=1}^{n} \Phi(\mathbf{s}_i) \right)
\end{equation}
where \( \Phi \) is a learned function (e.g., an MLP) applied independently to each element \( \mathbf{s}_i \in \mathcal{S} \), the summation (or an alternative permutation-invariant operator such as mean or max) aggregates the transformed elements, and \( \Psi \) is another learned function that maps the aggregated result to the final output. This framework serves as a universal approximator for functions over sets and forms the theoretical basis of the Deep Sets architecture.

In this work, the Deep Sets framework is adapted to estimate the modal weights \( \widehat{\Gamma}_\mathcal{M} \) from the dominant eigenvalue set \( \Lambda_\mathcal{M} \). Since the task is permutation equivariant, the architecture is extended accordingly, as described in the next section.

\subsection{Deep Sets-Inspired Model}
\begin{figure}[t]
\centering
\includegraphics[width=0.55\columnwidth]{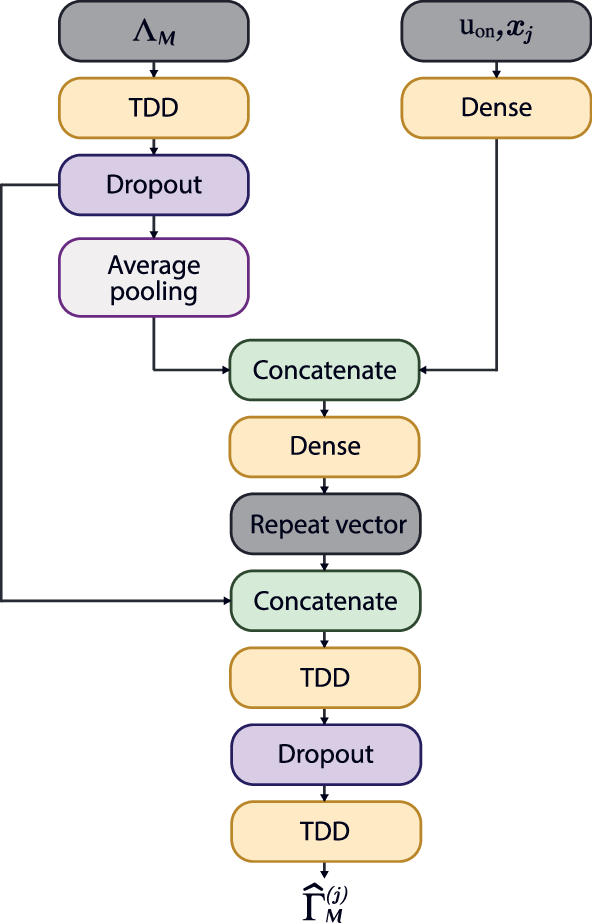}
\caption{DeepSet-Inspired Network}
\label{fig:DeepSet}
\end{figure}

The model is designed to be permutation equivariant with respect to the dominant eigenvalue set \( \Lambda_\mathcal{M} \), while conditioning on system-level features (\( \mathbf{u}_{\text{on}} \) and \( x_j \)). The complete model architecture is illustrated in Fig.~\ref{fig:DeepSet} and can be expressed as
\begin{align}
\label{eq:deep_set_ispired}
    \widehat{\Gamma}_\mathcal{M}^{(j)} &= \Psi\left( \phi(\lambda_i),\, c \right) , \quad \text{for all } \lambda_i \in \Lambda_\mathcal{M} \\
    c & = \rho\left( \sum_{\lambda_i \in \Lambda_\mathcal{M}} \Phi(\lambda_i),\, h(\mathbf{u}_{\text{on}}, x_j) \right)
\end{align}

\subsubsection{Input Representation}  
The input is divided into two components: eigenvalue-based features and flat features. Each eigenvalue is converted to its polar form \([|\lambda|, \angle\lambda]\) and stacked into a tensor of shape \([N, M, 2]\), where \( N \) is the batch size and \( M \) is the number of eigenvalues. The flat inputs \( \mathbf{u}_{\text{on}} \) and \( x_j \) are standardized to ensure consistent scaling across features.

\subsubsection{Deep Sets Block (Permutation-Invariant Branch)}  
The eigenvalue set \( \Lambda_\mathcal{M} \) is processed by applying a shared dense transformation \( \Phi(\cdot) \) independently to each element, mapping each eigenvalue into a \( d \)-dimensional latent space. In this work, \( \Phi(\cdot) \) is implemented using a \emph{TimeDistributed Dense} (TDD) layer from TensorFlow/Keras, which applies the same weights and biases across all modes. A dropout layer follows for regularization. This operation corresponds to the element-wise transformation term \( \Phi(\lambda_i) \) in Eq.~\ref{eq:deep_set_ispired}. The resulting latent representations are then aggregated using average pooling, producing a single \( d \)-dimensional vector that combines contributions from all modes equally—summation \( \sum_{\lambda_i \in \Lambda_\mathcal{M}} \Phi(\lambda_i) \) in Eq.~\ref{eq:deep_set_ispired}.

\subsubsection{Flat Feature Conditioning}  
In parallel, the flat inputs \( \mathbf{u}_{\text{on}} \) and \( x_j \) are passed through a dense layer \( h(\cdot) \), which projects them into a latent space of dimension \( d \). The pooled output from the Deep Sets branch and the flat input projection are concatenated into a single vector of size \( 2d \), referred to as the global context vector. This vector is then passed through an additional dense layer, corresponding to the function \( \rho(\cdot) \) in Eq.~\ref{eq:deep_set_ispired}.

\subsubsection{Equivariant Extension and Output}  
To preserve permutation equivariance, the global context vector \( c \) is repeated \( M \) times—where \( M \) is the number of eigenvalues in \( \Lambda_\mathcal{M} \)—and concatenated with each eigenvalue’s encoded representation \( \phi(\lambda_i) \). This ensures that each mode is enriched with the same global context while maintaining input order. The resulting tensor of shape \([N, M, 2d]\) is passed through another block of \emph{TimeDistributed Dense} (TDD) and dropout layers, enabling fusion of the global and local features. A final TDD layer then produces the predicted set of complex modal weights \( \widehat{\Gamma}_\mathcal{M}^{(j)} \). 

The function \( \Psi(\cdot) \) in Eq.~\ref{eq:deep_set_ispired} corresponds to the final sequence of TDD layers, which map the concatenated pair \([ \phi(\lambda_i), c ]\) to the polar embedding \([|\gamma_i|, \sin(\angle \gamma_i), \cos(\angle \gamma_i)]\). This representation avoids angular discontinuities and enables smooth training. Each modal coefficient \( \gamma_i \) is then reconstructed as \(\gamma_i = |\gamma_i| \cdot e^{j \angle \gamma_i}\), with \(\angle \gamma_i = \arctan2(\sin(\angle \gamma_i), \cos(\angle \gamma_i))\).

The number of TDD and dropout layers, the depth of the dense transformations, and the hidden dimension \( d \) are tunable hyperparameters that can be adjusted to control the model’s capacity and regularization behavior.

\subsection{Loss Function}
To improve numerical stability and learning performance, each complex modal coefficient \( \gamma_i \in \widehat{\Gamma}_\mathcal{M}^{(j)} \) is transformed into polar form, represented as the tuple \( (|\gamma_i|, \angle \gamma_i) \). This transformation addresses issues with small magnitudes and angular discontinuities near the origin, which were found to hinder direct learning from raw complex values. By standardizing \( |\gamma_i| \) and encoding \( \angle \gamma_i \) as \( [\sin(\angle \gamma_i), \cos(\angle \gamma_i)] \), discontinuities at \( \pm\pi \) are avoided, providing smooth and bounded inputs.

The total loss is defined as a weighted sum of the magnitude and angular components,
\begin{equation}
\label{eq:loss}
\mathcal{L}_{\text{polar}} = \alpha \cdot \mathcal{L}_r + \beta \cdot \mathcal{L}_\theta
\end{equation}
where \( \alpha \) and \( \beta \) are weighting coefficients that control the relative contribution of each term.

The magnitude loss \( \mathcal{L}_r \) is computed as the mean squared error between the predicted and true magnitudes:
\begin{equation}
\mathcal{L}_r = \frac{1}{n} \sum_{i=1}^{n} \left( |\gamma_i|^{\text{true}} - |\gamma_i|^{\text{pred}} \right)^2
\end{equation}

The angular loss \( \mathcal{L}_\theta \) is calculated using the vector difference between the predicted and true direction vectors,
\begin{equation}
\mathcal{L}_\theta = \frac{1}{n} \sum_{i=1}^{n} \left[ \cos^{-1} \left( \frac{\mathbf{v}_i^{\text{true}} \cdot \mathbf{v}_i^{\text{pred}}}{\|\mathbf{v}_i^{\text{true}}\| \cdot \|\mathbf{v}_i^{\text{pred}}\|} \right) \right]^2
\end{equation}
where \( n \) is the number of modes per sample, and \( \mathbf{v}_i = [\sin(\angle \gamma_i), \cos(\angle \gamma_i)] \) represents the angular direction vector of the predicted or true value. The argument of the inverse cosine is the normalized dot product between \( \mathbf{v}_i^{\text{true}} \) and \( \mathbf{v}_i^{\text{pred}} \), which computes the cosine of the angular difference between them.

\subsection{Data Preparation}
\label{sec:datapreparation}
Since frequency control modes are largely insensitive to variations in load flow, topology, and dispatch, complex modal weights depend primarily on unit commitment, modal structure, and disturbance characteristics \cite{zelaya2025modal}. Therefore, the model is trained by varying the unit commitment and disturbances, without requiring data from multiple operating conditions. This is an advantage over traditional deep learning approaches that must span a wide range of operating points for generalization. In contrast, the modal-based formulation enables predictions grounded in system dynamics, as the eigenstructure remains fixed and only disturbance-specific coefficients vary. Algorithm~\ref{alg:data_collection} outlines the data generation process for generator trip scenarios.

\begin{algorithm}[H]
\caption{Data Collection}
\label{alg:data_collection}
\begin{algorithmic}[1]

\State \textbf{Input:} Set of unit commitments \( \mathcal{U} \), scaling factors \( \boldsymbol{\alpha} \)
\State \textbf{Output:} Dataset \( \mathcal{D} = \left\{ \left( \Lambda_{\mathcal{M}}, \mathbf{u}_{\text{on}}, i, \Delta P_L^{(i)}, \Gamma_{\mathcal{M}}^{(i)} \right) \right\} \)

\ForAll{unit commitment \( \mathbf{u}_{\text{on}} \in \mathcal{U} \)}
    \State Linearize the system and compute modal set \( \Lambda_{\mathcal{M}} \)
    
    \ForAll{generator \( i \in \mathbf{u}_{\text{on}} \)}
        \ForAll{scaling factor \( \alpha_i \in \boldsymbol{\alpha} \)}
            \State Scale the power output of generator \( i \) by \( \alpha_i \)
            \State Trip generator \( i \)
            \State Compute power imbalance \( \Delta P_L^{(i)} \)
            \State Compute the set of modal weights \( \Gamma_{\mathcal{M}}^{(i)} \)
            \State \( \mathcal{D} \gets \mathcal{D} \cup \left\{ \left( \Lambda_{\mathcal{M}}, \mathbf{u}_{\text{on}}, i, \Delta P_L^{(i)}, \Gamma_{\mathcal{M}}^{(i)} \right) \right\} \)
        \EndFor
    \EndFor
\EndFor
\end{algorithmic}
\end{algorithm}

For each predefined unit commitment \( \mathbf{u}_{\text{on}} \in \mathcal{U} \), the system is linearized and the set of frequency control modes \( \Lambda_{\mathcal{M}} \) is computed. Note that this is done just for one operating condition, not multiple. 

Subsequently, if only generator trips are considered as disturbances, each committed generator is subjected to a series of trip scenarios by scaling its output using a predefined set of factors \( \boldsymbol{\alpha} \). These factors are chosen to span the generator’s operating range between \( P_{\text{min}} \) and \( P_{\text{max}} \), with each scaled output representing a distinct contingency case. For each scenario, the resulting active power imbalance \( \Delta P_L^{(i)} \) is recorded, and the complex modal weights \( \Gamma_i \) are computed using Eq.~\eqref{eq:modal_coefficients}. While the modal structure \( \Lambda_{\mathcal{M}} \) remains constant for a given unit commitment, the weights \( \Gamma_i \) vary with the disturbance magnitude.

The resulting dataset \( \mathcal{D} \) consists of tuples \((\Lambda_{\mathcal{M}}, \mathbf{u}_{\text{on}}, i, \Delta P_L^{(i)}, \Gamma_i)\), capturing the relationship between the system’s modal structure, the generator trip event, the resulting power imbalance, and the modal response.

\section{Case Study}
This section evaluates the performance of the proposed framework using two benchmark test systems: the IEEE 10-machine 39-bus system and the IEEE 118-bus system. The method is also compared against XGBoost and a feedforward neural network framework. All simulations are performed in MATLAB using the Power System Simulator (PSSim)~\cite{MartinezLizana2025}, a tool developed in-house.

\subsection{IEEE 10-Machine 39-Bus System}
This system consists of 10 SGs modeled using two-axis machine representations, IEEE Type-1 exciters, and three types of turbine governors: TGOV1, IEESGO, and GAST. System parameters are based on~\cite{chow2020power}, with governor settings adopted from~\cite{neplan2015turbine}. The full data set is available in~\cite{MartinezLizana2025}.

A total of 9,700 stable samples were generated using 155 unique unit commitment configurations. Generator trip events were simulated by scaling the power output of each generator using factors in steps of \( \alpha \in [0.35:0.05:0.95] \). For simplicity, only generator trips were considered in this study; however, load steps could also be incorporated into the training process. The dataset was split 70/30 for training and validation. The DeepSets-inspired model and framework were implemented in Python 3.10.15 using TensorFlow/Keras and Scikit-learn.

To evaluate the final framework, a set of dispatch scenarios and their corresponding unit commitments were generated by solving a simplified economic dispatch problem. Quadratic cost functions were assigned to each generator, with coefficients randomly varied within a small range around their nominal values. Table~\ref{tab:gen_output_scenarios} presents five representative test scenarios used to evaluate generalization performance.

\begin{table}[h!]
\centering
\scriptsize
\caption{\\ Test Scenarios: Generator Dispatch (MW)}
\label{tab:gen_output_scenarios}
\begin{tabular}{lccccc}
\toprule
\textbf{Gen} & \textbf{S1} & \textbf{S2} & \textbf{S3} & \textbf{S4} & \textbf{S5} \\
\midrule
G1  & 400.77 & 174.48 & 239.51 & 70.75  & 253.72 \\
G2  & 646.00 & 406.81 & 415.23 & 646.00 & 323.59 \\
G3  & 118.95 & 0      & 229.59 & 187.60 & 0      \\
G4  & 386.01 & 652.00 & 323.17 & 652.00 & 219.19 \\
G5  & 84.79  & 0      & 0      & 0      & 67.58  \\
G6  & 687.00 & 687.00 & 687.00 & 687.00 & 687.00 \\
G7  & 0      & 78.40  & 0      & 0      & 160.12 \\
G8  & 450.82 & 200.59 & 347.93 & 0      & 0      \\
G9  & 59.77  & 211.72 & 677.71 & 575.33 & 665.88 \\
G10 & 865.89 & 788.99 & 679.87 & 781.32 & 722.92 \\
\midrule
Demand & 3700 & 3200 & 3600 & 3600 & 3100 \\
\bottomrule
\end{tabular}
\end{table}

\begin{figure}[t!]
    \centering
    \subfloat[]{%
        \includegraphics[width=0.8\columnwidth]{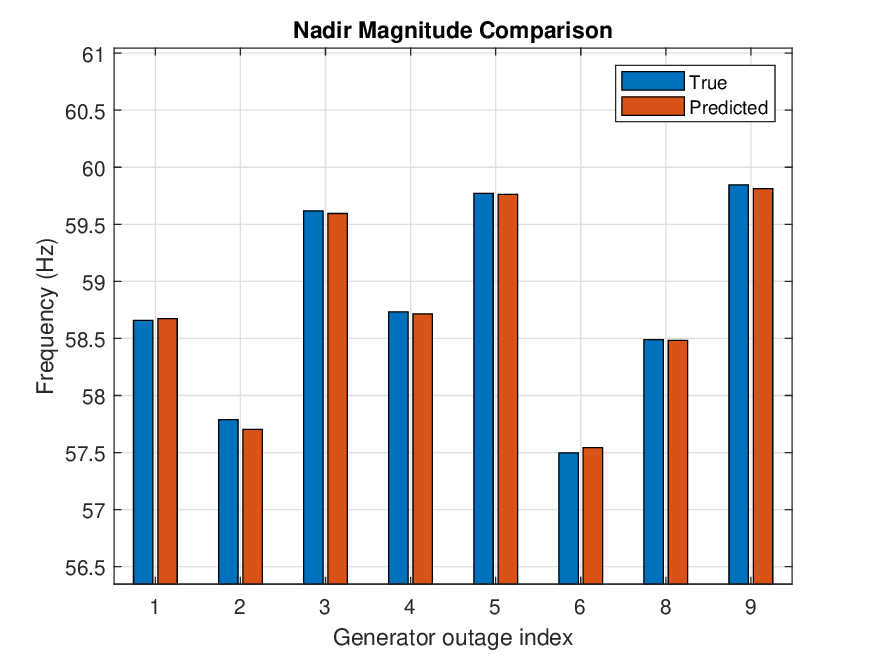}
        \label{fig:nadir_magnitude}
    }\\[0.1cm]
    \subfloat[]{%
        \includegraphics[width=0.8\columnwidth]{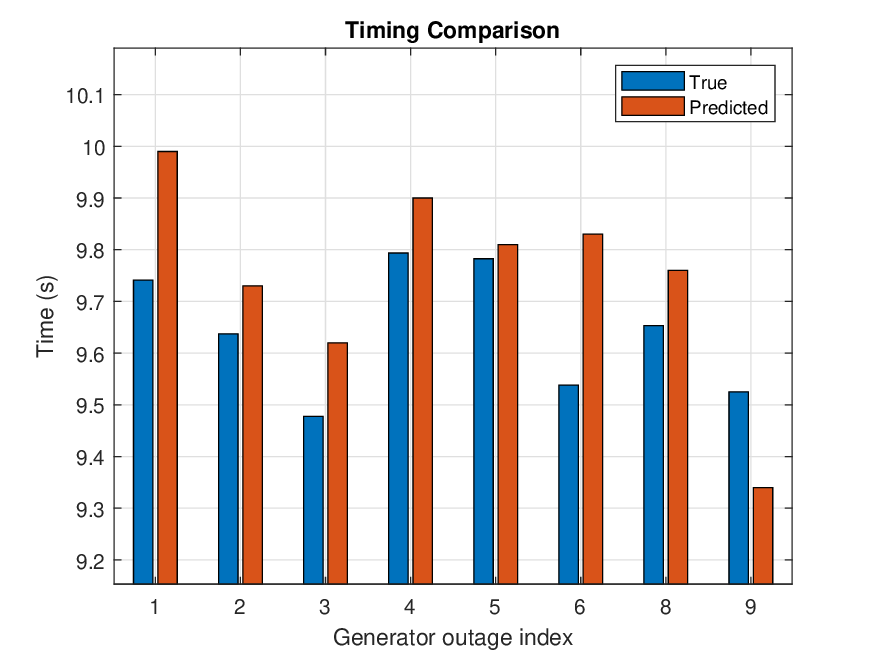}
        \label{fig:nadir_timing}
    }
    \caption{Comparison of actual vs. predicted results for 39 bus system: (a) Nadir Magnitude, (b) Nadir Timing}
    \label{fig:nadir_comparison_NE}
\end{figure}

Figure~\ref{fig:nadir_comparison_NE} illustrates the prediction results for Scenario 1 (S1). The framework processes all eight possible generator trips in 0.65~s, achieving a Mean Absolute Error (MAE) of 0.02~Hz for nadir magnitude and 0.15~s for timing. The corresponding Mean Absolute Percentage Errors (MAPE) are 0.05\% and 1.56\%, respectively. Although the timing error is slightly higher, it remains well within acceptable bounds—especially when compared with reported MAPE values in the literature, which range from 7.69\% to 19.97\% \cite{liu2020analytical,egido2009maximum}.

Table~\ref{tab:performance_metrics} summarizes the results across five test scenarios. The proposed framework achieves an average computation time of 0.678~s per case, highlighting its ability to deliver fast, accurate predictions without full system simulation. Low average errors in nadir magnitude and timing confirm strong generalization to unseen dispatch conditions. By relying only on dispatch and unit commitment data, the framework enhances situational awareness and supports real-time or near-real-time operational decision-making through rapid frequency nadir prediction.

\begin{table}[h!]
\centering
\scriptsize
\caption{\\Performance Metrics per Scenario}
\label{tab:performance_metrics}
\begin{tabular}{lccccc|c}
\toprule
 & \textbf{S1} & \textbf{S2} & \textbf{S3} & \textbf{S4} & \textbf{S5} & \textbf{Avg.} \\
\midrule
Nadir MAE (Hz)        & 0.029 & 0.015 & 0.015 & 0.038 & 0.016 & 0.023 \\
Timing MAE (s)        & 0.155 & 0.200 & 0.133 & 0.680 & 0.107 & 0.255 \\
Time per batch (s)    & 0.69  & 0.65  & 0.72  & 0.66  & 0.67  & 0.678 \\
\bottomrule
\end{tabular}
\end{table}

An additional case study is presented in which a $20\%$ inverter-based resource (IBR) penetration is incorporated. This is represented by four IBRs modeled using the renewable energy generator type-B (REGC-B), along with the controllers described in~\cite{ramasubramanian2016converter}. This model, recommended by EPRI~\cite{ramasubramanian2023}, is commonly used in commercial software to assess the impact of renewable integration on grid performance and reliability. The full dataset is available in~\cite{MartinezLizana2025}. Using the same settings as the previous case and training the model with 12{,}900 stable scenarios, the system is evaluated under a different set of dispatch scenarios and unit commitments. Table~\ref{tab:gen_output_scenarios2} presents five representative cases. The corresponding performance metrics are shown in Table~\ref{tab:performance_metrics2}, demonstrating the robustness of the method even when IBRs are incorporated.

\begin{table}[h!]
\centering
\scriptsize
\caption{\\ Test Scenarios: Generator Dispatch (MW) including IBRs}
\label{tab:gen_output_scenarios2}
\begin{tabular}{lccccc}
\toprule
\textbf{Gen} & \textbf{S1} & \textbf{S2} & \textbf{S3} & \textbf{S5} & \textbf{S7} \\
\midrule
SG1   & 367.39 & 410.01 & 109.68 & 214.82 & 0 \\
SG2   & 0      & 592.01 & 646.00 & 221.68 & 221.32 \\
SG3   & 0      & 0      & 0      & 0      & 68.05 \\
SG4   & 652.00 & 652.00 & 652.00 & 0      & 383.57 \\
SG5   & 111.66 & 115.97 & 0      & 0      & 0 \\
SG6   & 687.00 & 687.00 & 687.00 & 687.00 & 663.85 \\
SG7   & 0      & 0      & 0      & 189.97 & 0 \\
SG8   & 129.90 & 74.60  & 0      & 301.78 & 225.97 \\
SG9   & 477.69 & 113.03 & 480.05 & 673.16 & 211.83 \\
SG10  & 630.74 & 733.49 & 803.38 & 526.68 & 477.49 \\
IBR1  & 184.91 & 204.38 & 204.38 & 170.31 & 136.25 \\
IBR2  & 237.74 & 262.77 & 262.77 & 218.97 & 175.18 \\
IBR3  & 158.50 & 175.18 & 175.18 & 145.98 & 116.79 \\
IBR4  & 158.50 & 175.18 & 175.18 & 145.98 & 116.79 \\
\midrule
Demand & 3800 & 4200 & 4200 & 3500 & 2800 \\
\bottomrule
\end{tabular}
\end{table}

\begin{table}[h!]
\centering
\scriptsize
\caption{\\Performance Metrics per Scenario including IBRs}
\label{tab:performance_metrics2}
\begin{tabular}{lccccc|c}
\toprule
 & \textbf{S1} & \textbf{S2} & \textbf{S3} & \textbf{S4} & \textbf{S5} & \textbf{Avg.} \\
\midrule
Nadir MAE (Hz)        & 0.028  & 0.022  & 0.028  & 0.014  & 0.021  & 0.0226 \\
Timing MAE (s)        & 0.2953 & 0.300  & 0.360  & 0.220  & 0.460  & 0.3271 \\
Performance Score     & 0.711  & 0.683  & 0.673  & 0.663  & 0.761  & 0.698 \\
\bottomrule
\end{tabular}
\end{table}

\subsection{IEEE 118-Bus System}
The system consists of 118 buses and 19 SGs, modeled using two-axis machine representations, IEEE Type-1 exciters, and a mix of TGOV1, IEEESGO, and GAST governors. The dataset was generated using 40 unit commitment configurations with \( \alpha \in [0.35{:}0.05{:}0.95] \), combined with generator trip disturbances, resulting in a total of 8,600 stable samples. As in the previous case, the dataset was split 70/30 for training and validation.

\begin{figure}[t!]
    \centering
    \subfloat[]{%
        \includegraphics[width=0.8\columnwidth]{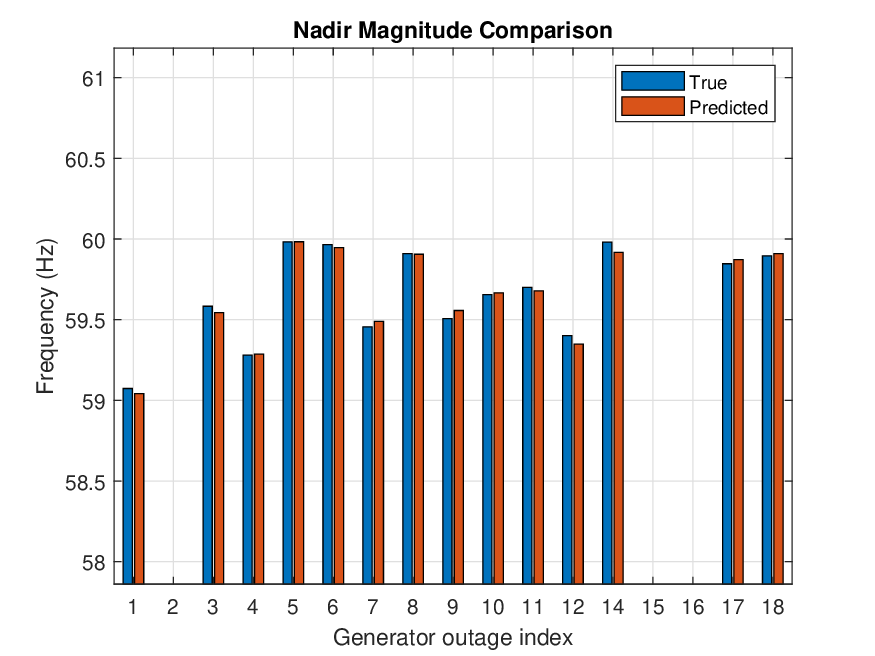}
        \label{fig:nadir_magnitude}
    }\\[0.1cm]
    \subfloat[]{%
        \includegraphics[width=0.8\columnwidth]{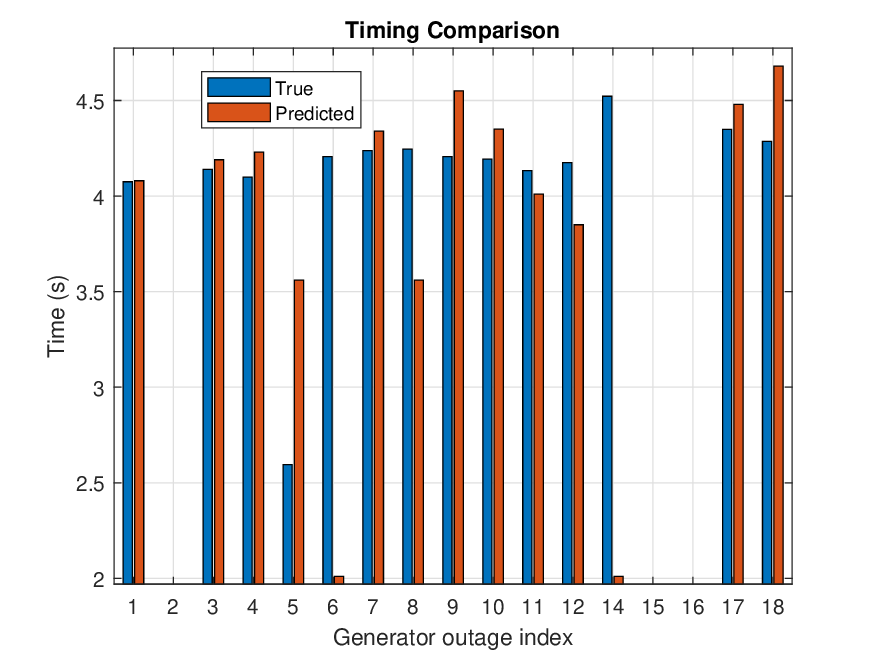}
        \label{fig:nadir_timing}
    }
    \caption{Comparison of actual vs. predicted results for the IEEE 118-bus system: (a) Nadir Magnitude, (b) Nadir Timing}
    \label{fig:nadir_comparison_118}
\end{figure}

The dispatch of a representative case study is shown in Table~\ref{tab:gen_output_single}, and Fig.~\ref{fig:nadir_comparison_118} compares predicted and actual nadir values. A total of 14 generator trips were evaluated in 1.15~s. Trips at G2, G15, G16, and G19 led to unstable conditions and non-convergent time-domain simulations. For the remaining stable cases, the model achieved a mean absolute error (MAE) of 0.026~Hz and 0.58~s, with mean absolute percentage errors (MAPE) of 0.044\% for magnitude and 14.49\% for timing. The higher timing error is primarily due to two outlier cases (6 and 14), which significantly deviated from the true timing values. Despite this, magnitude prediction remained highly accurate.

\begin{table}[h!]
\centering
\scriptsize
\caption{\\Generator Dispatch (MW)}
\label{tab:gen_output_single}
\setlength{\tabcolsep}{3pt}
\begin{tabular}{*{10}{c}}
\toprule
\textbf{G1} & \textbf{G2} & \textbf{G3} & \textbf{G4} & \textbf{G5} & \textbf{G6} & \textbf{G7} & \textbf{G8} & \textbf{G9} & \textbf{G10} \\
\midrule
550 & 185 & 320 & 414 & 7 & 19 & 296 & 48 & 255 & 203 \\
\toprule
\textbf{G11} & \textbf{G12} & \textbf{G13} & \textbf{G14} & \textbf{G15} & \textbf{G16} & \textbf{G17} & \textbf{G18} & \textbf{G19} & \\
\midrule
188 & 332 & 0 & 4 & 363 & 136 & 40 & 36 & 805 & \\
\bottomrule
\end{tabular}
\end{table}

\subsection{Comparison with purely data-driven methods}
To benchmark against purely data-driven methods, the proposed framework is adapted to operate without any model-based components. Instead, it uses direct grid measurements—assumed to be available from PMUs—to predict \( f_{\text{nadir}} \) and \( t_{\text{nadir}} \) through regression. Although the disturbance list is generated in the same way as in the proposed framework, this variant bypasses modal-based computations entirely. The input vector for this formulation is:
\begin{equation}
    \mathbf{x} = [ \mathbf{V}, \boldsymbol{\theta}, \text{Dispatch}, \Delta P_{L}, \mathbf{u}_{\text{on}}, i_{\text{trip}} ]
\end{equation}
where \( \mathbf{V} \in \mathbb{R}^{n_b} \) and \( \boldsymbol{\theta} \in \mathbb{R}^{n_b} \) are the bus voltage magnitudes and voltage angles, respectively; \textit{Dispatch} \( \in \mathbb{R}^{n_g} \) denotes the generator real power outputs; \( \Delta P_{L} \in \mathbb{R} \) is the disturbance magnitude; \( \mathbf{u}_{\text{on}} \in \{0,1\}^{n_g} \) is the binary unit commitment vector indicating the on/off status of each generator; and \( i_{\text{trip}} \) is the index of the tripped generator.

XGBoost \cite{chen2016xgboost,chen2019xgboost} and FNN \cite{burkov2019hundred} are trained using the same training cases as the DeepSets-inspired model, with input features adjusted accordingly—100 for the IEEE 39-bus and 276 for the IEEE 118-bus system.

Table~\ref{tab:combined_model_comparison_mae} compares training and evaluation performance. In this context, training refers to the 30\% test split used during model development, while evaluation error corresponds to newly generated scenarios based on random dispatch and unit commitment—i.e., a dataset with a different distribution. While both the FNN and XGBoost perform well on the training data, their evaluation errors increase significantly, particularly for nadir magnitude, reaching up to 0.27~Hz and 0.35~Hz, respectively. Such deviations could compromise protection schemes like under-frequency load shedding. In contrast, the DeepSets-inspired model exhibits lower and more consistent evaluation errors across both systems. Note that training error is not reported for this model, as it does not directly predict the frequency nadir and timing but instead estimates the modal coefficient set \( \Gamma_{\mathcal{M}}^{(j)} \).

Although FNN and XGBoost performance could be enhanced by incorporating additional operating scenarios, richer measurement sets, or more complex AI architectures, these improvements would demand greater computational resources and infrastructure—especially widespread PMU availability. In contrast, the proposed hybrid framework demonstrates stronger generalization capabilities by leveraging system dynamics through the modal decomposition. Rather than learning the final output in isolation, it captures the underlying structure of the system response. Moreover, it only requires generator power output information, making it suitable for real-world deployment without the need for full-state observability.

Regarding time performance, all analyses were conducted on a personal computer equipped with an Intel(R) Core(TM) i7-9750H CPU, NVIDIA GeForce GTX 1650 GPU, and 16~GB of RAM. XGBoost was the fastest, requiring 0.07~s on average to evaluate all trip cases, followed by FFNN at 1.13~s. The DeepSets-inspired model required 1.76~s on average due to additional modal-based processing but achieved higher accuracy, stronger generalization, and required fewer measurement inputs. 
\begin{table*}[t]
\centering
\scriptsize
\caption{\\ Error comparison for the 39-bus and 118-bus systems}
\label{tab:combined_model_comparison_mae}
\setlength{\tabcolsep}{4pt}
\begin{tabular}{llcc|cc|c|cc|cc|c}
\toprule
& & \multicolumn{5}{c|}{\textbf{39-Bus System}} & \multicolumn{5}{c}{\textbf{118-Bus System}} \\
& & \multicolumn{2}{c|}{\textbf{FFNN}} & \multicolumn{2}{c|}{\textbf{XGBoost}} & \textbf{DeepSet} 
  & \multicolumn{2}{c|}{\textbf{FFNN}} & \multicolumn{2}{c|}{\textbf{XGBoost}} & \textbf{DeepSet} \\
\cmidrule(r){3-4} \cmidrule(r){5-6} \cmidrule(r){7-7} \cmidrule(r){8-9} \cmidrule(r){10-11} \cmidrule(r){12-12}
\textbf{Metric} & \textbf{Unit} 
& Train & Eval & Train & Eval & Eval 
& Train & Eval & Train & Eval & Eval \\
\midrule
Nadir MAE & (Hz) 
& 0.048 & 0.27 & 0.024 & 0.32 & \textbf{0.023} 
& 0.015 & 0.35 & 0.007 & 0.11 & \textbf{0.026} \\
Timing MAE & (s) 
& 0.081 & 0.80 & 0.098 & 2.32 & \textbf{0.25} 
& 0.048 & 0.78 & 0.060 & 0.55 & \textbf{0.58} \\
\bottomrule
\end{tabular}
\end{table*}

\section{Conclusion}
This paper introduced a hybrid data-model framework for frequency dynamic security assessment in power systems. Building on the modal-based system frequency response and nadir prediction approach, the method employs a DeepSets-inspired neural network to estimate complex modal coefficients that characterize the system’s response to disturbances. By replacing repeated model-based computations with a trained estimator, the framework enables fast and accurate predictions of frequency nadir and its timing across varying operating conditions. Validation on the IEEE 39-bus and 118-bus systems demonstrates strong generalization and low prediction error compared to conventional machine learning models such as XGBoost and feedforward neural networks, which require full system measurements. While the DeepSets-inspired model incurs slightly higher computation time due to modal reconstruction, it significantly outperforms baseline models in accuracy and robustness, using fewer input features. These results highlight the potential of hybrid approaches that combine physical insights with learning-based parameter estimation to provide scalable, efficient tools for enhancing situational awareness and supporting real-time decision-making.

\appendices

\bibliographystyle{IEEEtran}
\bibliography{Biblio}



\end{document}